\documentclass[reprint,
 amsmath,amssymb,
 aps,
]{revtex4-2}

\usepackage[english]{babel}
\usepackage[T1]{fontenc}
\usepackage{amsmath,amssymb,amsfonts}
\usepackage{graphicx}
\usepackage{bm}
\usepackage{hyperref}
\usepackage{physics}
\usepackage{float}

\begin{document}

\title{Classical and single photon memory devices based on polariton lasers}

\author{D. Novokreschenov}
\affiliation{Abrikosov Center for Theoretical Physics, MIPT, Dolgoprudnyi, Moscow Region 141701, Russia}
\affiliation{Russian Quantum Center, Skolkovo, Moscow, 121205, Russia}
\author{A. Kudlis}
\affiliation{Science Institute, University of Iceland, Dunhagi-3, IS-107 Reykjavik, Iceland}
% \affiliation{Russian Quantum Center, Skolkovo, Moscow, 121205, Russia}
\author{A. V. Kavokin}
\affiliation{Abrikosov Center for Theoretical Physics, MIPT, Dolgoprudnyi, Moscow Region 141701, Russia}
\affiliation{Russian Quantum Center, Skolkovo, Moscow, 121205, Russia}
\affiliation{Department of Physics, St. Petersburg State University, University Embankment, 7/9, St. Petersburg, 199034, Russia}
\affiliation{School of Science, Westlake University, 18 Shilongshan Road, Hangzhou 310024, Zhejiang Province, China}

\date{\today}

\begin{abstract}
    Stimulated scattering of incoherent excitons into an exciton-polariton mode leads to the build-up of a polariton condensate whose polarization is sensitive to a small seeded population that triggers the stimulated process. We show, within a semiclassical stochastic Gross-Pitaevskii model, that this mechanism enables a robust polarization memory operation: the condensate tends to align its Stokes vector with that of the seed and to maintain it for times far exceeding an individual polariton lifetime. Importantly, this \textit{single-photon-seeded} regime is modeled as an initial weak excitation of the condensate mode. We quantify the memory performance by a classical polarization-alignment metric and find that the seed polarization can remain well preserved on a nanosecond timescale under realistic parameters.
\end{abstract}

\maketitle

%%%%%%%%%%%%%%%%%%%%%%%%%%%%%%%%%%%%%%%%%%%%%%%%%%
% \section{Introduction}
\textit{Introduction}---
Exciton-polaritons in microcavities are hybrid light-matter quasiparticles emerging from the strong coupling between quantum well excitons and cavity photons first observed by Weisbuch \textit{et al.}~\cite{WeisbuchPRL1992}. Polariton lasers are optical devices that take advantage of the coherence of Bose-Einstein condensates of exciton-polaritons to produce coherent and monochromatic light whose polarization properties may be controlled externally~\cite{DengRMP2010,KasprzakNature2006,BaliliScience2007,YamamotoScience2002}. These condensates belong to the class of driven--dissipative quantum fluids of light, where pumping and loss govern the steady state and yield excitation spectra distinct from equilibrium condensates~\cite{WoutersCarusottoPRL2007,CarusottoCiutiRMP2013}. Polariton lasers are based on semiconductor microcavities operating in the regime of strong exciton-photon coupling~\cite{WeisbuchPRL1992}. Importantly, polariton condensation and lasing can be realized from cryogenic to room temperature by choosing appropriate materials, allowing practical device concepts~\cite{ChristopoulosPRL2007,BaumbergPRL2008}. Non-resonant optical pumping or electronic injection leads to the formation of an incoherent exciton reservoir that feeds a coherent bosonic condensate of exciton-polaritons due to the stimulated inelastic exciton scattering~\cite{WoutersCarusottoPRL2007,DangPRL1998,SavvidisPRL2000,CiutiPRB2000,WertzPhys2012}. At the same time, resonant parametric processes -- where two pumped polaritons scatter into signal and idler states -- provide low-threshold amplification and coherent control that underpin many polaritonic circuit elements~\cite{SavvidisPRL2000,CiutiPRB2000}.
\begin{figure}[t!]
\centering
\includegraphics[width=0.99\linewidth]{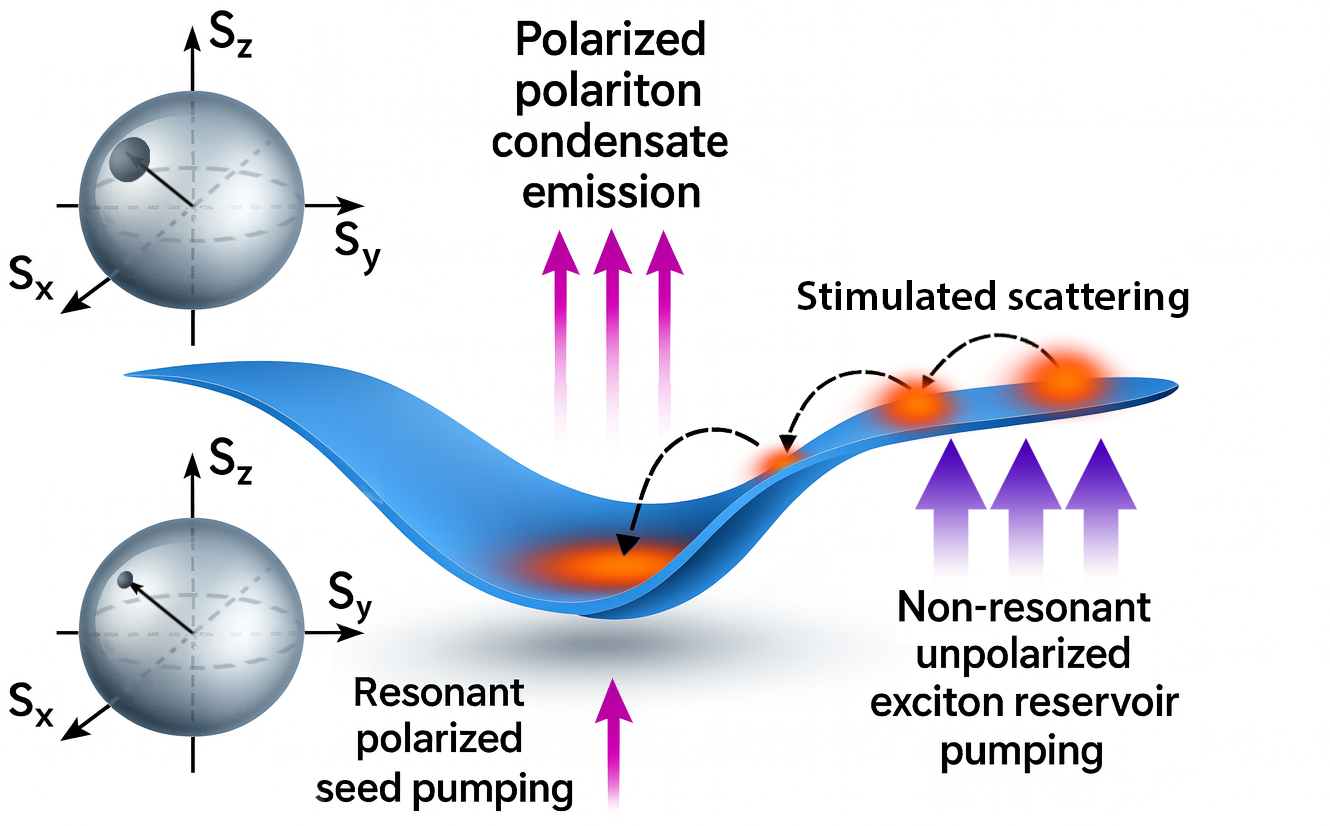} 
\caption{The pumping scheme of a polariton laser operating in the regime of optical memory. A polariton mode is seeded by resonant laser light. The seed polarization (shown at the bottom Poincar\'e sphere) is preserved and amplified due to the stimulated scattering of excitons from an incoherent exciton reservoir that is pumped non-resonantly. The polarization of light emitted by the condensate is governed by the seen polarization (upper Poincar\'e sphere).
\label{fig:dispersion}}
\end{figure}
\begin{figure*}
    \centering
\includegraphics[width=0.99\linewidth]{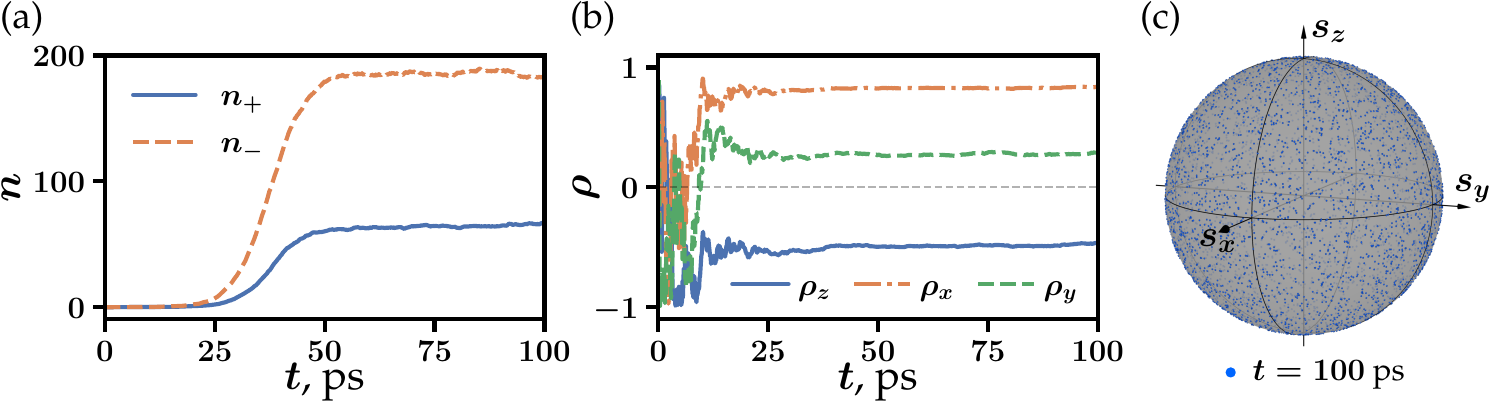}
\caption{\label{fig:no_seed}
Simulation for the zero initial condensate amplitude, $\psi(t{=}0)=0$ (no resonant seed). No structural anisotropy or external fields are included, so polarization is not pinned. (a) Time evolution of the circular-component occupations $n_\pm=\lvert\psi_\pm\rvert^2$ (solid: $\sigma^+$, dashed: $\sigma^-$) for a representative noise realization, showing condensate build-up due to saturable gain and approach to a steady state. 
(b) Degrees of polarization $\rho_i=2S_i/n$ ($i=x,y,z$) for the same trajectory: large stochastic swings at early times give way to a quasi-stationary Stokes vector selected randomly by fluctuations. 
(c) Ensemble distribution of the normalized Stokes vector $\mathbf{s}=2\mathbf{S}/n$ on the Poincar\'e sphere at delay $\tau=100$ ps, obtained from 5000 independent noise realizations; the nearly uniform coverage demonstrates that, without a seed, the final polarization is random.}
\end{figure*}
Amplification of the population of a polariton mode due to the stimulated scattering stabilizes spin and polarization of exciton-polaritons in the polariton lasing regime~\cite{BaumbergPRL2008,ReadPRB2009,ShelykhPRL2006,KasprzakPRB2007,LevratPRL2010,AdradosPRL2011}. In particular, the spinor nature of polaritons (two circular components, $\sigma^\pm$) allows a condensate to spontaneously select and maintain a polarization state, which can be pinned by anisotropies or, conversely, stochastically varied in the presence of noise~\cite{KasprzakPRB2007,LevratPRL2010,ReadPRB2009}. As we show in the present work, the polarization build-up effect can be exploited to create robust optical memory devices, where a polarization state can be preserved over characteristic times that exceed a single polariton lifetime by several orders of magnitude~\cite{GippiusPRL2007,LiewPRL2008}. Beyond fundamental interest, these mechanisms directly connect to all-optical information processing: experiment and theory now demonstrate polariton neurons, logic gates, and spin switches that use polarization as a binary degree of freedom~\cite{LiewPRL2008,LiNP2024,ZhaoNC2024,OpalaOME2023}. In particular, ultra-fast temporal logic gates operating at room temperature and picosecond-scale spin switching showcase the feasibility of polarization logic and latch-like memory devices based on semiconductor microcavities~\cite{LiNP2024,ZhaoNC2024}.

The formation of a polariton condensate can be considered as an (out of equilibrium) phase transition~\cite{BaumbergPRL2008}. Within this concept, the build-up of a bosonic condensate can be characterized by the emergence of an order parameter $\psi_\sigma\equiv\expval{\hat a_\sigma}$, where $\hat a_\sigma$ is the polariton annihilation operator corresponding to the polarization $\sigma$. Due to the stochastic nature of the scattering process, the relative phase of the order parameters for different polarizations is chosen randomly, leading to a random polarization of the resulting condensate~\cite{LaussyPRB2006,ReadPRB2009}. In the driven--dissipative case, fluctuations of the exciton reservoir and quantum noise seed initial state of the condensate. This seeding can bias polarization selection, which is accounted for if modeling the system with stochastic equations for $\psi_\sigma$ and, when needed, for the reservoir~\cite{WoutersCarusottoPRL2007,vanKampen2007}. Stochastic models correctly describe the experimental data on polarization pinning, multistability, and spontaneous switching ~\cite{KasprzakPRB2007,LevratPRL2010}.

It is important to note, that because of bosonic amplification, the scattering probability for excitons having the same polarization with the condensate is enhanced proportionally to the occupation number of the condensate. If, initially, the polariton mode is resonantly populated with a small seed of exciton-polaritons, a correlation between the final polarization of the condensate and the initial polarization of the seed can be anticipated~\cite{RuboPRL2003,RuboPSSA2004,WoutersPRB2007}.

In this Letter, we develop a formalism to describe the formation and polarization of a polariton condensate formed by scattering of excitons from an incoherent reservoir triggered by a small seed of polaritons introduced by resonant optical pumping. We solve a stochastic Gross–Pitaevskii equation for the condensate order parameter and compute the polarization of the emitted light~\cite{MandelWolf1995,vanKampen2007} as a function of the time elapsed after the seeding. Throughout, the single-photon–seeded case is treated as an initial weak excitation of the condensate mode within a semiclassical description. Accordingly, our results demonstrate a quantum-initiated, but classical polarization memory operation~\cite{Note1}.
\begin{figure*}
    \centering
    \includegraphics[width=0.99\linewidth]{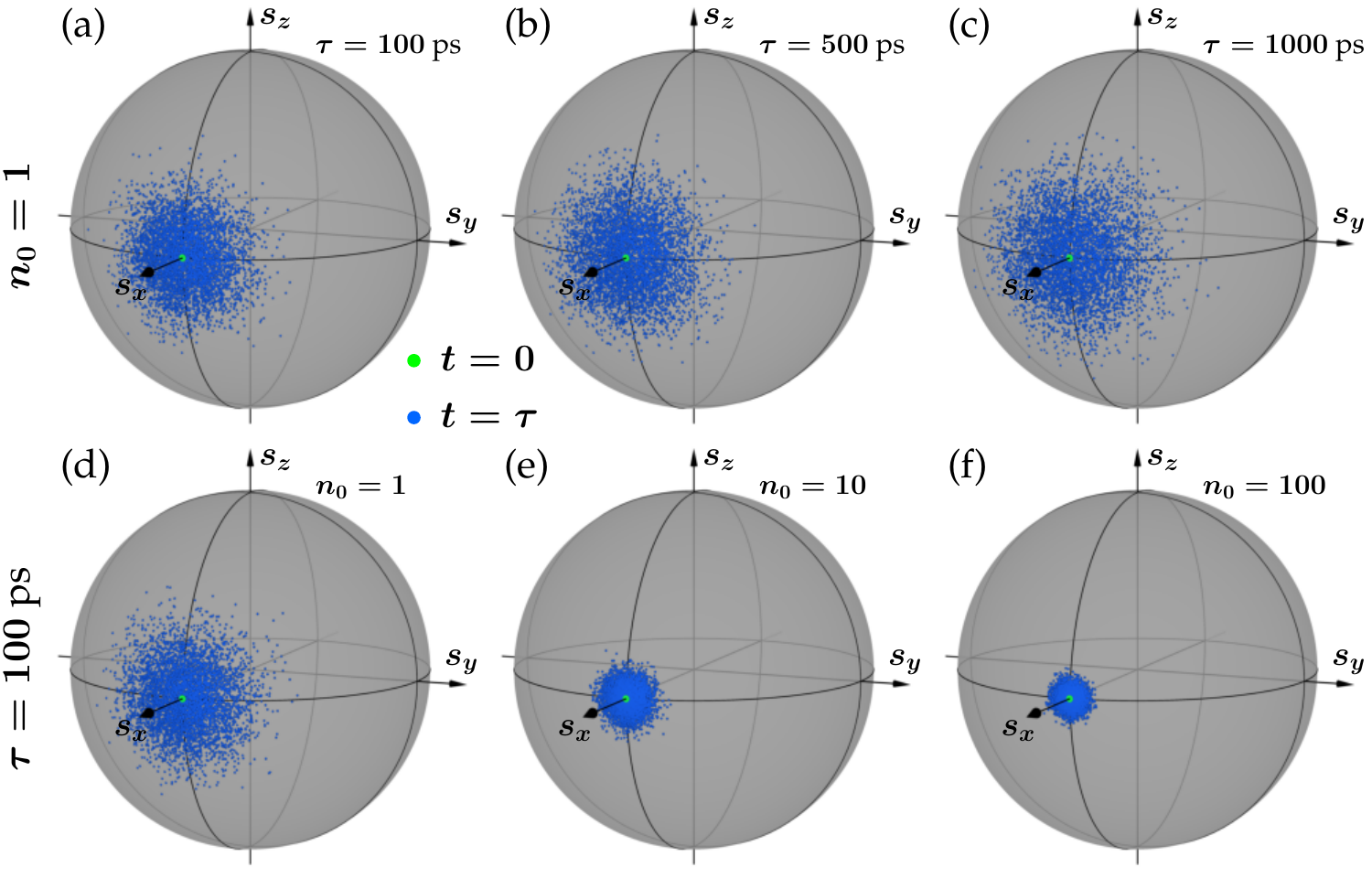}
\caption{\label{fig:distr_diff_tau}
Ensemble distributions of the normalized Stokes vector $\mathbf{s}$ on the Poincar\'e sphere, computed over 5000 independent noise realizations. 
(a–c) Variation with the observation delay $\tau$ (100, 500, and 1000 ps, respectively) for a single–polariton seed ($n_0=1$): the cloud of outcomes remains centered near the seed polarization but broadens with time, indicating gradual loss of alignment in the absence of pinning.  (d–f) Dependence on the initial seed population $n_0$ (1, 10, and 100, respectively) at fixed delay $\tau=100$ ps: the distribution tightens around the seed as $n_0$ increases, reflecting stronger bosonic stimulation and a faster condensate build-up. 
All simulations use a horizontally polarized seed with $\mathbf{s}_0=\{1,0,0\}$. The green point indicates the initial Stokes vector $\mathbf{s}_0$, while blue points show $\mathbf{s}$ at time $\tau$; axes $s_x$, $s_y$, and $s_z$ are labeled in each panel.}
\end{figure*}
\allowdisplaybreaks
% \section{Model and Formalism}

\textit{Model and formalism}---We shall consider the formation of a spinor condensate of exciton-polaritons in a planar semiconductor microcavity. We assume that the condensate is being formed in the lowest energy state of the lower polariton dispersion branch (see Fig~\ref{fig:dispersion}) which is frequently the case, experimentally. The formation of a polariton condensate is manifested by the buildup of an order parameter $\psi_\sigma = \expval{\hat a_{0,\sigma}}$, which defines the coherence properties of the condensate. We use a stochastic Gross-Pitaevskii equation to describe the evolution of the order parameter.

We describe the dynamics of the stimulated scattering of excitons to the condensate by $\hbar\partial_t \psi = W_\text{in}\psi$. A similar term, $\hbar\partial_t \psi = -\hbar\gamma\psi$, describes the decay of particles; unless otherwise stated we set $\gamma = 0.5~\mathrm{ps^{-1}}$. The interplay of gain and loss makes the dynamics of the condensate unstable: if the gain rate exceeds the loss rate, the condensate grows indefinitely; if the loss rate exceeds the gain rate, the condensate vanishes. In practice, for nonresonantly pumped solid-state systems, the gain is saturable. The saturation of gain can be accounted for by introduction of a density-dependent gain rate, $W_\text{in}=P-\Gamma\abs{\psi}^2$, which helps bringing the system to a stationary regime at $\abs{\psi}^2=(P-\hbar\gamma)/\Gamma$. Combining the gain and loss terms with the stochastic noise, we obtain the following equation:

\begin{equation}
i\hbar\pdv{\psi_\sigma}{t}=\frac i2\qty[P-\Gamma\qty(\abs{\psi_\sigma}^2+\abs{\psi_{\overline\sigma}}^2)-\hbar\gamma]\psi_\sigma + \hbar\xi,
\end{equation}
where $\sigma$ denotes one of two circular polarizations and $\xi$ is the Gaussian white noise whose dispersion is proportional to the polariton income rate:
\begin{align}
\expval{\xi(t)\xi^*(t')} &\propto \frac12 W_\text{in}(t)\delta(t-t') \\ \nonumber\ &= \frac14 \qty[P-\Gamma\qty(\abs{\psi_\sigma}^2+\abs{\psi_{\overline\sigma}}^2)]\delta(t-t').
\end{align}

The Stokes vector of light emitted by a polariton condensate is linked to its order parameter as follows:
\begin{align}
&S_x = \frac12 \qty(\psi_{-1}^*\psi_{+1} + \psi_{+1}^*\psi_{-1}), \\
&S_y = \frac i2 \qty(\psi_{-1}^*\psi_{+1} - \psi_{+1}^*\psi_{-1}), \\
&S_z = \frac12 \qty(\abs{\psi_{+1}}^2 - \abs{\psi_{-1}}^2).
\end{align}

To find the polarization of the emitted light, it is sufficient to compute the normalized Stokes vector
\begin{equation}
\vb{s}\equiv \frac{2\vb S}{n},
\end{equation}
where $n=\abs{\psi_+}^2+\abs{\psi_-}^2$ is the total occupation number of the condensate. The components of this unit vector define a point on the surface of the Poincaré sphere.

% \section{Numerical simulations}
\textit{Numerical simulations}---As a reference point, we consider the case of no initial seed characterized by $\psi_{t=0}=0$. From the beginning, we should emphasize that no structural anisotropy or external fields are used to pin polarization; alignment arises from bosonic stimulation conditioned by the seeded pseudospin. Panels (a) and (b) in Fig.~\ref{fig:no_seed} present time evolutions of the occupation numbers $n_\sigma=\abs{\psi_\sigma}^2$ and the degree of polarization $\rho_i=2S_i/n$ during the condensate formation for an arbitrary realization of noise. As can be seen from these simulations, during the formation of the condensate its polarization fluctuates randomly, while the polarization formed by the end of the condensate formation is preserved for a long time. In the absence of initial seed, the final polarization of the condensate is random: it is distributed uniformly over the surface of the Poincaré sphere, as shown in panel (c). Note that the model assumes no polarization pinning caused by structural anisotropy or external fields here.

\begin{figure}
    \centering
    \includegraphics[width=0.99\linewidth]{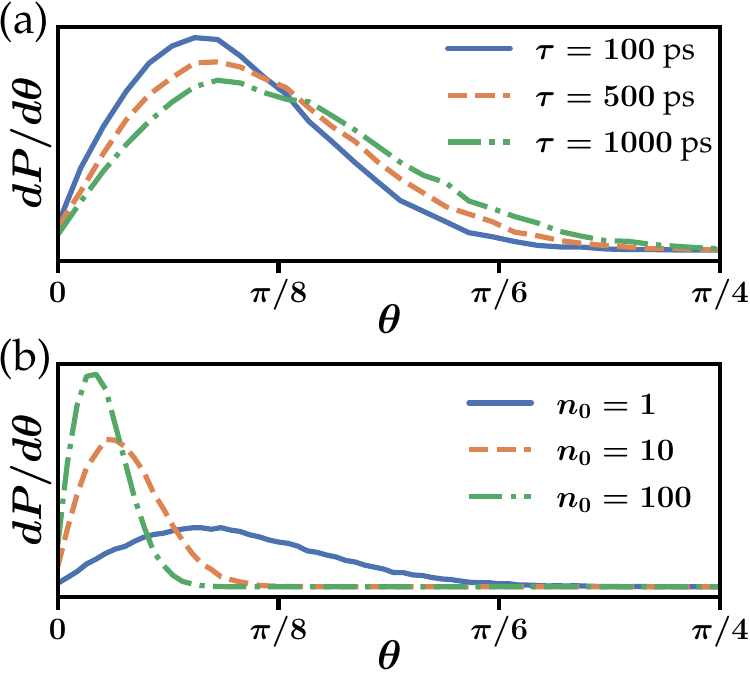}
    \caption{\label{fig:prob_density_diff_tau} The probability densities of deviations of the Stokes vector of the polariton condensate from the initial seed Stokes vector by an angle $\theta$: (a) For the time delays of $\tau = 100$ ps, $\tau = 500$ ps and $\tau = 1000$ ps and single polariton initial seed ($n_0=1$); (b) For initial seed populations of $n_0=1$, $n_0=10$ and $n_0=100$ polaritons and time delay $\tau=100$ ps.}
\end{figure}
\begin{figure}[b!]
    \centering
    \includegraphics[width=0.99\linewidth]{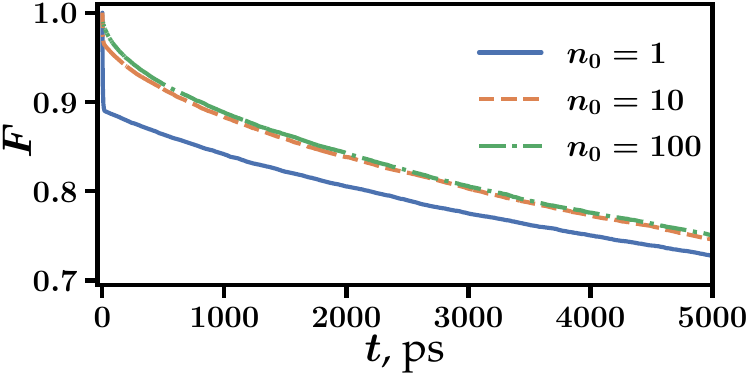}
    \caption{\label{fig:F}
Time dependence of the polarization-alignment metric (PAM) $F(t)=1-2\langle\theta\rangle/\pi$ for different seed occupation numbers $n_{0}$, obtained by averaging over 5000 stochastic realizations. By definition, $F=1$ corresponds to perfect alignment of the condensate Stokes vector with the seed, while $F=0$ indicates a random final polarization. Larger seeds ($n_{0}=10,100$) start from higher $F$ and retain alignment longer (slower decay), reflecting stronger bosonic stimulation and a shorter build-up time. The single-polariton case ($n_{0}=1$) shows the weakest retention.}
\end{figure}

In contrast, if a polariton condensate is built upon an initial seed of polarized polaritons, its polarization is no more random. Clearly, because of the bosonic stimulation of scattering, the polarization of seed-polaritons may be maintained and even amplified. 
Consequently, the distribution of the resulting polarizations of the polariton condensates is no longer uniform. It peaks around the seed's polarization, as shown in Fig.~\ref{fig:distr_diff_tau}. For these simulations we use an initial seed characterized by a horizontal polarization and a normalized Stokes vector $\vb s_0 = \qty{1, 0, 0}$. We also consider different time delays (panels (a)-(c)) and different seed occupation numbers (panels (d)-(f)). In the single-photon-seeded case ($n_0=1$) we model the seed as an initial weak excitation of the condensate mode within a semiclassical description. As a measure of the deviation between the condensate polarization and the initial seed polarization, we use the angle $\theta$ between their Stokes vectors:
\begin{equation}
\theta = \arccos(\vb s\cdot\vb s_0).
\end{equation}
The probability density of the deviation of the condensate polarization with respect to the seed polarization in the case of a single initial polariton is shown in Fig.~\ref{fig:prob_density_diff_tau}(a) as a function of angle $\theta$. Due to stochastic scattering processes, the condensate's polarization still may vary, which is why the distribution function broadens. Over time, the distribution will tend to the uniform one, as in Fig.~\ref{fig:no_seed}(c).

The spread of the Stokes vector around the initial state depends not only on time but also on the population of the initial seed. The simulation results in Fig.~\ref{fig:distr_diff_tau}(d)-(f) and Fig.~\ref{fig:prob_density_diff_tau}(b) show that the deviation of the polarization from its initial value is smaller for larger seed populations. This is related to the even greater bosonic enhancement as compared to the small seed limit. The shorter condensate formation time also contributes to the more pronounced memory effect in the case of a large seed.

To numerically estimate the spread of polarization with respect to its initial value, one can use the average angle of deviation of the Stokes vector, which ranges from $\expval{\theta}=\pi/2$ in the case of a random final polarization to $\expval{\theta}=0$ in the case of perfectly preserved polarization.  We then define a polarization-alignment metric (PAM)
\begin{equation}
F(t)\equiv1-\frac{2\expval{\theta}}{\pi},
\end{equation}
which ranges from $0$ (random) to $1$ (perfect alignment) in the corresponding cases. This is a classical figure of merit and should not be confused with the quantum-state fidelity used in quantum-initiated polarization memory benchmarks. The time dependence of PAM for different initial seeds is shown in Fig.~\ref{fig:F}.
These simulations demonstrate that the polarization of the initial seed is preserved with good accuracy for hundreds of picoseconds, even for a single initial polariton. This time exceeds the characteristic polariton lifetime in planar microcavities by two orders of magnitude.

% \section{Conclusion}
\textit{Conclusion}---We have demonstrated that exciton-polariton condensates can act as robust optical memories for polarization states. Bosonic stimulation of the exciton scattering to the polariton mode during the condensate formation ensures that the condensate is likely to keep the same polarization as the initial seed of polaritons introduced by resonant optical pumping. 
%Numerical simulations show that this polarization memory effect persists for hundreds of picoseconds which exceeds by far the single polariton lifetime.

Remarkably, the memory effect persists in the single-photon–seeded regime: in our semiclassical model, a single-photon seed can imprint a polarization that remains aligned for nanosecond-scale delays, as quantified by $F(t)$. For larger seed populations (large $n_0$), our simulations yield a polarization-alignment metric of $F(\tau)\simeq 0.9$ at the same delay. Moreover, it can be further enhanced if larger seed populations are introduced. We conclude that bosonic stimulation of exciton scattering in semiconductor microcavities may be used for the realization of classical and quantum optical memory devices. Finally, we notice that the use of high-quality factor microcavities ~\cite{BruneCommMat2025} and neuromorphic architectures ~\cite{SedovLight2025,OpalaOME2023} pave the way to the realization of an information processing platform that converts, preserves, and processes data all-optically ~\cite{CuevasSciAdv2018}. In this context, proposed polariton memory devices are expected to find a significant application area.

\textit{Acknowledgment}--AVK acknowledges support from Saint Petersburg State University (Research Grant No. 125022803069-4) and from the Innovation Program for Quantum Science and Technology (No. 2021ZD0302704). The work of AK is supported by the Icelandic Research Fund (Ranns\'oknasj\'o{\dh}ur, Grant No.~2410550).

\bibliography{lit}   

\begin{thebibliography}{33}%
\makeatletter
\providecommand \@ifxundefined [1]{%
 \@ifx{#1\undefined}
}%
\providecommand \@ifnum [1]{%
 \ifnum #1\expandafter \@firstoftwo
 \else \expandafter \@secondoftwo
 \fi
}%
\providecommand \@ifx [1]{%
 \ifx #1\expandafter \@firstoftwo
 \else \expandafter \@secondoftwo
 \fi
}%
\providecommand \natexlab [1]{#1}%
\providecommand \enquote  [1]{``#1''}%
\providecommand \bibnamefont  [1]{#1}%
\providecommand \bibfnamefont [1]{#1}%
\providecommand \citenamefont [1]{#1}%
\providecommand \href@noop [0]{\@secondoftwo}%
\providecommand \href [0]{\begingroup \@sanitize@url \@href}%
\providecommand \@href[1]{\@@startlink{#1}\@@href}%
\providecommand \@@href[1]{\endgroup#1\@@endlink}%
\providecommand \@sanitize@url [0]{\catcode `\\12\catcode `\$12\catcode `\&12\catcode `\#12\catcode `\^12\catcode `\_12\catcode `\%12\relax}%
\providecommand \@@startlink[1]{}%
\providecommand \@@endlink[0]{}%
\providecommand \url  [0]{\begingroup\@sanitize@url \@url }%
\providecommand \@url [1]{\endgroup\@href {#1}{\urlprefix }}%
\providecommand \urlprefix  [0]{URL }%
\providecommand \Eprint [0]{\href }%
\providecommand \doibase [0]{https://doi.org/}%
\providecommand \selectlanguage [0]{\@gobble}%
\providecommand \bibinfo  [0]{\@secondoftwo}%
\providecommand \bibfield  [0]{\@secondoftwo}%
\providecommand \translation [1]{[#1]}%
\providecommand \BibitemOpen [0]{}%
\providecommand \bibitemStop [0]{}%
\providecommand \bibitemNoStop [0]{.\EOS\space}%
\providecommand \EOS [0]{\spacefactor3000\relax}%
\providecommand \BibitemShut  [1]{\csname bibitem#1\endcsname}%
\let\auto@bib@innerbib\@empty
%</preamble>
\bibitem [{\citenamefont {Weisbuch}\ \emph {et~al.}(1992)\citenamefont {Weisbuch}, \citenamefont {Nishioka}, \citenamefont {Ishikawa},\ and\ \citenamefont {Arakawa}}]{WeisbuchPRL1992}%
  \BibitemOpen
  \bibfield  {author} {\bibinfo {author} {\bibfnamefont {C.}~\bibnamefont {Weisbuch}}, \bibinfo {author} {\bibfnamefont {M.}~\bibnamefont {Nishioka}}, \bibinfo {author} {\bibfnamefont {A.}~\bibnamefont {Ishikawa}},\ and\ \bibinfo {author} {\bibfnamefont {Y.}~\bibnamefont {Arakawa}},\ }\bibfield  {title} {\bibinfo {title} {Observation of the coupled exciton-photon mode splitting in a semiconductor microcavity},\ }\href {https://doi.org/10.1103/PhysRevLett.69.3314} {\bibfield  {journal} {\bibinfo  {journal} {Phys. Rev. Lett.}\ }\textbf {\bibinfo {volume} {69}},\ \bibinfo {pages} {3314} (\bibinfo {year} {1992})}\BibitemShut {NoStop}%
\bibitem [{\citenamefont {Deng}\ \emph {et~al.}(2010)\citenamefont {Deng}, \citenamefont {Haug},\ and\ \citenamefont {Yamamoto}}]{DengRMP2010}%
  \BibitemOpen
  \bibfield  {author} {\bibinfo {author} {\bibfnamefont {H.}~\bibnamefont {Deng}}, \bibinfo {author} {\bibfnamefont {H.}~\bibnamefont {Haug}},\ and\ \bibinfo {author} {\bibfnamefont {Y.}~\bibnamefont {Yamamoto}},\ }\bibfield  {title} {\bibinfo {title} {Exciton-polariton bose-einstein condensation},\ }\href {https://doi.org/10.1103/RevModPhys.82.1489} {\bibfield  {journal} {\bibinfo  {journal} {Rev. Mod. Phys.}\ }\textbf {\bibinfo {volume} {82}},\ \bibinfo {pages} {1489} (\bibinfo {year} {2010})}\BibitemShut {NoStop}%
\bibitem [{\citenamefont {Kasprzak}\ \emph {et~al.}(2006)\citenamefont {Kasprzak}, \citenamefont {Richard}, \citenamefont {Kundermann}, \citenamefont {Baas}, \citenamefont {Jeambrun}, \citenamefont {Keeling}, \citenamefont {Marchetti}, \citenamefont {Szyma{\'n}ska}, \citenamefont {Andr{\'e}}, \citenamefont {Staehli}, \citenamefont {Savona}, \citenamefont {Littlewood}, \citenamefont {Deveaud},\ and\ \citenamefont {Dang}}]{KasprzakNature2006}%
  \BibitemOpen
  \bibfield  {author} {\bibinfo {author} {\bibfnamefont {J.}~\bibnamefont {Kasprzak}}, \bibinfo {author} {\bibfnamefont {M.}~\bibnamefont {Richard}}, \bibinfo {author} {\bibfnamefont {S.}~\bibnamefont {Kundermann}}, \bibinfo {author} {\bibfnamefont {A.}~\bibnamefont {Baas}}, \bibinfo {author} {\bibfnamefont {P.}~\bibnamefont {Jeambrun}}, \bibinfo {author} {\bibfnamefont {J.~M.~J.}\ \bibnamefont {Keeling}}, \bibinfo {author} {\bibfnamefont {F.~M.}\ \bibnamefont {Marchetti}}, \bibinfo {author} {\bibfnamefont {M.~H.}\ \bibnamefont {Szyma{\'n}ska}}, \bibinfo {author} {\bibfnamefont {R.}~\bibnamefont {Andr{\'e}}}, \bibinfo {author} {\bibfnamefont {J.~L.}\ \bibnamefont {Staehli}}, \bibinfo {author} {\bibfnamefont {V.}~\bibnamefont {Savona}}, \bibinfo {author} {\bibfnamefont {P.~B.}\ \bibnamefont {Littlewood}}, \bibinfo {author} {\bibfnamefont {B.}~\bibnamefont {Deveaud}},\ and\ \bibinfo {author} {\bibfnamefont {L.~S.}\ \bibnamefont {Dang}},\ }\bibfield  {title} {\bibinfo {title} {Bose--einstein condensation of
  exciton polaritons},\ }\href {https://doi.org/10.1038/nature05131} {\bibfield  {journal} {\bibinfo  {journal} {Nature}\ }\textbf {\bibinfo {volume} {443}},\ \bibinfo {pages} {409} (\bibinfo {year} {2006})}\BibitemShut {NoStop}%
\bibitem [{\citenamefont {Balili}\ \emph {et~al.}(2007)\citenamefont {Balili}, \citenamefont {Hartwell}, \citenamefont {Snoke}, \citenamefont {Pfeiffer},\ and\ \citenamefont {West}}]{BaliliScience2007}%
  \BibitemOpen
  \bibfield  {author} {\bibinfo {author} {\bibfnamefont {R.}~\bibnamefont {Balili}}, \bibinfo {author} {\bibfnamefont {V.}~\bibnamefont {Hartwell}}, \bibinfo {author} {\bibfnamefont {D.}~\bibnamefont {Snoke}}, \bibinfo {author} {\bibfnamefont {L.}~\bibnamefont {Pfeiffer}},\ and\ \bibinfo {author} {\bibfnamefont {K.}~\bibnamefont {West}},\ }\bibfield  {title} {\bibinfo {title} {Bose-einstein condensation of microcavity polaritons in a trap},\ }\href {https://doi.org/10.1126/science.1140990} {\bibfield  {journal} {\bibinfo  {journal} {Science}\ }\textbf {\bibinfo {volume} {316}},\ \bibinfo {pages} {1007} (\bibinfo {year} {2007})}\BibitemShut {NoStop}%
\bibitem [{\citenamefont {Deng}\ \emph {et~al.}(2002)\citenamefont {Deng}, \citenamefont {Weihs}, \citenamefont {Santori}, \citenamefont {Bloch},\ and\ \citenamefont {Yamamoto}}]{YamamotoScience2002}%
  \BibitemOpen
  \bibfield  {author} {\bibinfo {author} {\bibfnamefont {H.}~\bibnamefont {Deng}}, \bibinfo {author} {\bibfnamefont {G.}~\bibnamefont {Weihs}}, \bibinfo {author} {\bibfnamefont {C.}~\bibnamefont {Santori}}, \bibinfo {author} {\bibfnamefont {J.}~\bibnamefont {Bloch}},\ and\ \bibinfo {author} {\bibfnamefont {Y.}~\bibnamefont {Yamamoto}},\ }\bibfield  {title} {\bibinfo {title} {Condensation of semiconductor microcavity exciton polaritons},\ }\href {https://doi.org/10.1126/science.1074464} {\bibfield  {journal} {\bibinfo  {journal} {Science}\ }\textbf {\bibinfo {volume} {298}},\ \bibinfo {pages} {199} (\bibinfo {year} {2002})}\BibitemShut {NoStop}%
\bibitem [{\citenamefont {Wouters}\ and\ \citenamefont {Carusotto}(2007)}]{WoutersCarusottoPRL2007}%
  \BibitemOpen
  \bibfield  {author} {\bibinfo {author} {\bibfnamefont {M.}~\bibnamefont {Wouters}}\ and\ \bibinfo {author} {\bibfnamefont {I.}~\bibnamefont {Carusotto}},\ }\bibfield  {title} {\bibinfo {title} {Excitations in a nonequilibrium bose-einstein condensate of exciton polaritons},\ }\href {https://doi.org/10.1103/PhysRevLett.99.140402} {\bibfield  {journal} {\bibinfo  {journal} {Phys. Rev. Lett.}\ }\textbf {\bibinfo {volume} {99}},\ \bibinfo {pages} {140402} (\bibinfo {year} {2007})}\BibitemShut {NoStop}%
\bibitem [{\citenamefont {Carusotto}\ and\ \citenamefont {Ciuti}(2013)}]{CarusottoCiutiRMP2013}%
  \BibitemOpen
  \bibfield  {author} {\bibinfo {author} {\bibfnamefont {I.}~\bibnamefont {Carusotto}}\ and\ \bibinfo {author} {\bibfnamefont {C.}~\bibnamefont {Ciuti}},\ }\bibfield  {title} {\bibinfo {title} {Quantum fluids of light},\ }\href {https://doi.org/10.1103/RevModPhys.85.299} {\bibfield  {journal} {\bibinfo  {journal} {Rev. Mod. Phys.}\ }\textbf {\bibinfo {volume} {85}},\ \bibinfo {pages} {299} (\bibinfo {year} {2013})}\BibitemShut {NoStop}%
\bibitem [{\citenamefont {Christopoulos}\ \emph {et~al.}(2007)\citenamefont {Christopoulos}, \citenamefont {von H{\"o}gersthal}, \citenamefont {Grundy}, \citenamefont {Lagoudakis}, \citenamefont {Kavokin}, \citenamefont {Baumberg}, \citenamefont {Christmann}, \citenamefont {Butt{\'e}}, \citenamefont {Feltin}, \citenamefont {Carlin},\ and\ \citenamefont {Grandjean}}]{ChristopoulosPRL2007}%
  \BibitemOpen
  \bibfield  {author} {\bibinfo {author} {\bibfnamefont {S.}~\bibnamefont {Christopoulos}}, \bibinfo {author} {\bibfnamefont {G.~B.~H.}\ \bibnamefont {von H{\"o}gersthal}}, \bibinfo {author} {\bibfnamefont {A.~J.~D.}\ \bibnamefont {Grundy}}, \bibinfo {author} {\bibfnamefont {P.~G.}\ \bibnamefont {Lagoudakis}}, \bibinfo {author} {\bibfnamefont {A.~V.}\ \bibnamefont {Kavokin}}, \bibinfo {author} {\bibfnamefont {J.~J.}\ \bibnamefont {Baumberg}}, \bibinfo {author} {\bibfnamefont {G.}~\bibnamefont {Christmann}}, \bibinfo {author} {\bibfnamefont {R.}~\bibnamefont {Butt{\'e}}}, \bibinfo {author} {\bibfnamefont {E.}~\bibnamefont {Feltin}}, \bibinfo {author} {\bibfnamefont {J.-F.}\ \bibnamefont {Carlin}},\ and\ \bibinfo {author} {\bibfnamefont {N.}~\bibnamefont {Grandjean}},\ }\bibfield  {title} {\bibinfo {title} {Room-temperature polariton lasing in semiconductor microcavities},\ }\href {https://doi.org/10.1103/PhysRevLett.98.126405} {\bibfield  {journal} {\bibinfo  {journal} {Phys. Rev. Lett.}\ }\textbf {\bibinfo
  {volume} {98}},\ \bibinfo {pages} {126405} (\bibinfo {year} {2007})}\BibitemShut {NoStop}%
\bibitem [{\citenamefont {Baumberg}\ \emph {et~al.}(2008)\citenamefont {Baumberg}, \citenamefont {Kavokin}, \citenamefont {Christopoulos}, \citenamefont {Grundy}, \citenamefont {Butt{\'e}}, \citenamefont {Christmann}, \citenamefont {Solnyshkov}, \citenamefont {Malpuech}, \citenamefont {von H{\"o}gersthal}, \citenamefont {Feltin}, \citenamefont {Carlin},\ and\ \citenamefont {Grandjean}}]{BaumbergPRL2008}%
  \BibitemOpen
  \bibfield  {author} {\bibinfo {author} {\bibfnamefont {J.~J.}\ \bibnamefont {Baumberg}}, \bibinfo {author} {\bibfnamefont {A.~V.}\ \bibnamefont {Kavokin}}, \bibinfo {author} {\bibfnamefont {S.}~\bibnamefont {Christopoulos}}, \bibinfo {author} {\bibfnamefont {A.~J.~D.}\ \bibnamefont {Grundy}}, \bibinfo {author} {\bibfnamefont {R.}~\bibnamefont {Butt{\'e}}}, \bibinfo {author} {\bibfnamefont {G.}~\bibnamefont {Christmann}}, \bibinfo {author} {\bibfnamefont {D.~D.}\ \bibnamefont {Solnyshkov}}, \bibinfo {author} {\bibfnamefont {G.}~\bibnamefont {Malpuech}}, \bibinfo {author} {\bibfnamefont {G.~B.~H.}\ \bibnamefont {von H{\"o}gersthal}}, \bibinfo {author} {\bibfnamefont {E.}~\bibnamefont {Feltin}}, \bibinfo {author} {\bibfnamefont {J.-F.}\ \bibnamefont {Carlin}},\ and\ \bibinfo {author} {\bibfnamefont {N.}~\bibnamefont {Grandjean}},\ }\bibfield  {title} {\bibinfo {title} {Spontaneous polarization buildup in a room-temperature polariton laser},\ }\href {https://doi.org/10.1103/PhysRevLett.101.136409} {\bibfield
  {journal} {\bibinfo  {journal} {Phys. Rev. Lett.}\ }\textbf {\bibinfo {volume} {101}},\ \bibinfo {pages} {136409} (\bibinfo {year} {2008})}\BibitemShut {NoStop}%
\bibitem [{\citenamefont {Dang}\ \emph {et~al.}(1998)\citenamefont {Dang}, \citenamefont {Heger}, \citenamefont {Andr{\'e}}, \citenamefont {B{\oe}uf},\ and\ \citenamefont {Romestain}}]{DangPRL1998}%
  \BibitemOpen
  \bibfield  {author} {\bibinfo {author} {\bibfnamefont {L.~S.}\ \bibnamefont {Dang}}, \bibinfo {author} {\bibfnamefont {D.}~\bibnamefont {Heger}}, \bibinfo {author} {\bibfnamefont {R.}~\bibnamefont {Andr{\'e}}}, \bibinfo {author} {\bibfnamefont {F.}~\bibnamefont {B{\oe}uf}},\ and\ \bibinfo {author} {\bibfnamefont {R.}~\bibnamefont {Romestain}},\ }\bibfield  {title} {\bibinfo {title} {Stimulation of microcavity-polariton photoluminescence in semiconductor microcavities},\ }\href {https://doi.org/10.1103/PhysRevLett.81.3920} {\bibfield  {journal} {\bibinfo  {journal} {Phys. Rev. Lett.}\ }\textbf {\bibinfo {volume} {81}},\ \bibinfo {pages} {3920} (\bibinfo {year} {1998})}\BibitemShut {NoStop}%
\bibitem [{\citenamefont {Savvidis}\ \emph {et~al.}(2000)\citenamefont {Savvidis}, \citenamefont {Baumberg}, \citenamefont {Stevenson}, \citenamefont {Skolnick}, \citenamefont {Whittaker},\ and\ \citenamefont {Roberts}}]{SavvidisPRL2000}%
  \BibitemOpen
  \bibfield  {author} {\bibinfo {author} {\bibfnamefont {P.~G.}\ \bibnamefont {Savvidis}}, \bibinfo {author} {\bibfnamefont {J.~J.}\ \bibnamefont {Baumberg}}, \bibinfo {author} {\bibfnamefont {R.~M.}\ \bibnamefont {Stevenson}}, \bibinfo {author} {\bibfnamefont {M.~S.}\ \bibnamefont {Skolnick}}, \bibinfo {author} {\bibfnamefont {D.~M.}\ \bibnamefont {Whittaker}},\ and\ \bibinfo {author} {\bibfnamefont {J.~S.}\ \bibnamefont {Roberts}},\ }\bibfield  {title} {\bibinfo {title} {Angle-resonant stimulated polariton amplifier},\ }\href {https://doi.org/10.1103/PhysRevLett.84.1547} {\bibfield  {journal} {\bibinfo  {journal} {Phys. Rev. Lett.}\ }\textbf {\bibinfo {volume} {84}},\ \bibinfo {pages} {1547} (\bibinfo {year} {2000})}\BibitemShut {NoStop}%
\bibitem [{\citenamefont {Ciuti}\ \emph {et~al.}(2000)\citenamefont {Ciuti}, \citenamefont {Schwendimann}, \citenamefont {Deveaud},\ and\ \citenamefont {Quattropani}}]{CiutiPRB2000}%
  \BibitemOpen
  \bibfield  {author} {\bibinfo {author} {\bibfnamefont {C.}~\bibnamefont {Ciuti}}, \bibinfo {author} {\bibfnamefont {P.}~\bibnamefont {Schwendimann}}, \bibinfo {author} {\bibfnamefont {B.}~\bibnamefont {Deveaud}},\ and\ \bibinfo {author} {\bibfnamefont {A.}~\bibnamefont {Quattropani}},\ }\bibfield  {title} {\bibinfo {title} {Theory of the angle-resonant polariton amplifier},\ }\href {https://doi.org/10.1103/PhysRevB.62.R4825} {\bibfield  {journal} {\bibinfo  {journal} {Phys. Rev. B}\ }\textbf {\bibinfo {volume} {62}},\ \bibinfo {pages} {R4825} (\bibinfo {year} {2000})}\BibitemShut {NoStop}%
\bibitem [{\citenamefont {Wertz}\ \emph {et~al.}(2012)\citenamefont {Wertz}, \citenamefont {Amo}, \citenamefont {Solnyshkov}, \citenamefont {Ferrier}, \citenamefont {Liew}, \citenamefont {Sanvitto}, \citenamefont {Senellart}, \citenamefont {Sagnes}, \citenamefont {Lema\^{\i}tre}, \citenamefont {Kavokin}, \citenamefont {Malpuech},\ and\ \citenamefont {Bloch}}]{WertzPhys2012}%
  \BibitemOpen
  \bibfield  {author} {\bibinfo {author} {\bibfnamefont {E.}~\bibnamefont {Wertz}}, \bibinfo {author} {\bibfnamefont {A.}~\bibnamefont {Amo}}, \bibinfo {author} {\bibfnamefont {D.~D.}\ \bibnamefont {Solnyshkov}}, \bibinfo {author} {\bibfnamefont {L.}~\bibnamefont {Ferrier}}, \bibinfo {author} {\bibfnamefont {T.~C.~H.}\ \bibnamefont {Liew}}, \bibinfo {author} {\bibfnamefont {D.}~\bibnamefont {Sanvitto}}, \bibinfo {author} {\bibfnamefont {P.}~\bibnamefont {Senellart}}, \bibinfo {author} {\bibfnamefont {I.}~\bibnamefont {Sagnes}}, \bibinfo {author} {\bibfnamefont {A.}~\bibnamefont {Lema\^{\i}tre}}, \bibinfo {author} {\bibfnamefont {A.~V.}\ \bibnamefont {Kavokin}}, \bibinfo {author} {\bibfnamefont {G.}~\bibnamefont {Malpuech}},\ and\ \bibinfo {author} {\bibfnamefont {J.}~\bibnamefont {Bloch}},\ }\bibfield  {title} {\bibinfo {title} {Propagation and amplification dynamics of 1d polariton condensates},\ }\href {https://doi.org/10.1103/PhysRevLett.109.216404} {\bibfield  {journal} {\bibinfo  {journal} {Phys. Rev.
  Lett.}\ }\textbf {\bibinfo {volume} {109}},\ \bibinfo {pages} {216404} (\bibinfo {year} {2012})}\BibitemShut {NoStop}%
\bibitem [{\citenamefont {Read}\ \emph {et~al.}(2009)\citenamefont {Read}, \citenamefont {Liew}, \citenamefont {Rubo},\ and\ \citenamefont {Kavokin}}]{ReadPRB2009}%
  \BibitemOpen
  \bibfield  {author} {\bibinfo {author} {\bibfnamefont {D.}~\bibnamefont {Read}}, \bibinfo {author} {\bibfnamefont {T.~C.~H.}\ \bibnamefont {Liew}}, \bibinfo {author} {\bibfnamefont {Y.~G.}\ \bibnamefont {Rubo}},\ and\ \bibinfo {author} {\bibfnamefont {A.~V.}\ \bibnamefont {Kavokin}},\ }\bibfield  {title} {\bibinfo {title} {Stochastic polarization formation in exciton-polariton bose-einstein condensates},\ }\href {https://doi.org/10.1103/PhysRevB.80.195309} {\bibfield  {journal} {\bibinfo  {journal} {Phys. Rev. B}\ }\textbf {\bibinfo {volume} {80}},\ \bibinfo {pages} {195309} (\bibinfo {year} {2009})}\BibitemShut {NoStop}%
\bibitem [{\citenamefont {Shelykh}\ \emph {et~al.}(2006)\citenamefont {Shelykh}, \citenamefont {Rubo}, \citenamefont {Kavokin}, \citenamefont {Liew},\ and\ \citenamefont {Malpuech}}]{ShelykhPRL2006}%
  \BibitemOpen
  \bibfield  {author} {\bibinfo {author} {\bibfnamefont {I.~A.}\ \bibnamefont {Shelykh}}, \bibinfo {author} {\bibfnamefont {Y.~G.}\ \bibnamefont {Rubo}}, \bibinfo {author} {\bibfnamefont {A.~V.}\ \bibnamefont {Kavokin}}, \bibinfo {author} {\bibfnamefont {T.~C.~H.}\ \bibnamefont {Liew}},\ and\ \bibinfo {author} {\bibfnamefont {G.}~\bibnamefont {Malpuech}},\ }\bibfield  {title} {\bibinfo {title} {Polarization and propagation of exciton-polaritons in microcavities},\ }\href {https://doi.org/10.1103/PhysRevLett.97.066402} {\bibfield  {journal} {\bibinfo  {journal} {Phys. Rev. Lett.}\ }\textbf {\bibinfo {volume} {97}},\ \bibinfo {pages} {066402} (\bibinfo {year} {2006})}\BibitemShut {NoStop}%
\bibitem [{\citenamefont {Kasprzak}\ \emph {et~al.}(2007)\citenamefont {Kasprzak}, \citenamefont {Andr{\'e}}, \citenamefont {Dang}, \citenamefont {Shelykh}, \citenamefont {Kavokin}, \citenamefont {Rubo}, \citenamefont {Kavokin},\ and\ \citenamefont {Malpuech}}]{KasprzakPRB2007}%
  \BibitemOpen
  \bibfield  {author} {\bibinfo {author} {\bibfnamefont {J.}~\bibnamefont {Kasprzak}}, \bibinfo {author} {\bibfnamefont {R.}~\bibnamefont {Andr{\'e}}}, \bibinfo {author} {\bibfnamefont {L.~S.}\ \bibnamefont {Dang}}, \bibinfo {author} {\bibfnamefont {I.~A.}\ \bibnamefont {Shelykh}}, \bibinfo {author} {\bibfnamefont {A.~V.}\ \bibnamefont {Kavokin}}, \bibinfo {author} {\bibfnamefont {Y.~G.}\ \bibnamefont {Rubo}}, \bibinfo {author} {\bibfnamefont {K.~V.}\ \bibnamefont {Kavokin}},\ and\ \bibinfo {author} {\bibfnamefont {G.}~\bibnamefont {Malpuech}},\ }\bibfield  {title} {\bibinfo {title} {Build up and pinning of linear polarization in the bose condensates of exciton polaritons},\ }\href {https://doi.org/10.1103/PhysRevB.75.045326} {\bibfield  {journal} {\bibinfo  {journal} {Phys. Rev. B}\ }\textbf {\bibinfo {volume} {75}},\ \bibinfo {pages} {045326} (\bibinfo {year} {2007})}\BibitemShut {NoStop}%
\bibitem [{\citenamefont {Levrat}\ \emph {et~al.}(2010)\citenamefont {Levrat}, \citenamefont {Butt{\'e}}, \citenamefont {Christian}, \citenamefont {Glauser}, \citenamefont {Feltin}, \citenamefont {Carlin},\ and\ \citenamefont {Grandjean}}]{LevratPRL2010}%
  \BibitemOpen
  \bibfield  {author} {\bibinfo {author} {\bibfnamefont {J.}~\bibnamefont {Levrat}}, \bibinfo {author} {\bibfnamefont {R.}~\bibnamefont {Butt{\'e}}}, \bibinfo {author} {\bibfnamefont {T.}~\bibnamefont {Christian}}, \bibinfo {author} {\bibfnamefont {M.}~\bibnamefont {Glauser}}, \bibinfo {author} {\bibfnamefont {E.}~\bibnamefont {Feltin}}, \bibinfo {author} {\bibfnamefont {J.-F.}\ \bibnamefont {Carlin}},\ and\ \bibinfo {author} {\bibfnamefont {N.}~\bibnamefont {Grandjean}},\ }\bibfield  {title} {\bibinfo {title} {Pinning and depinning of the polarization of exciton-polariton condensates at room temperature},\ }\href {https://doi.org/10.1103/PhysRevLett.104.166402} {\bibfield  {journal} {\bibinfo  {journal} {Phys. Rev. Lett.}\ }\textbf {\bibinfo {volume} {104}},\ \bibinfo {pages} {166402} (\bibinfo {year} {2010})}\BibitemShut {NoStop}%
\bibitem [{\citenamefont {Adrados}\ \emph {et~al.}(2011)\citenamefont {Adrados}, \citenamefont {Liew}, \citenamefont {Amo}, \citenamefont {Martin}, \citenamefont {Sanvitto}, \citenamefont {Anton}, \citenamefont {Giacobino}, \citenamefont {Kavokin}, \citenamefont {Bramati},\ and\ \citenamefont {Vi{\~n}a}}]{AdradosPRL2011}%
  \BibitemOpen
  \bibfield  {author} {\bibinfo {author} {\bibfnamefont {C.}~\bibnamefont {Adrados}}, \bibinfo {author} {\bibfnamefont {T.~C.~H.}\ \bibnamefont {Liew}}, \bibinfo {author} {\bibfnamefont {A.}~\bibnamefont {Amo}}, \bibinfo {author} {\bibfnamefont {M.~D.}\ \bibnamefont {Martin}}, \bibinfo {author} {\bibfnamefont {D.}~\bibnamefont {Sanvitto}}, \bibinfo {author} {\bibfnamefont {C.}~\bibnamefont {Anton}}, \bibinfo {author} {\bibfnamefont {E.}~\bibnamefont {Giacobino}}, \bibinfo {author} {\bibfnamefont {A.}~\bibnamefont {Kavokin}}, \bibinfo {author} {\bibfnamefont {A.}~\bibnamefont {Bramati}},\ and\ \bibinfo {author} {\bibfnamefont {L.}~\bibnamefont {Vi{\~n}a}},\ }\bibfield  {title} {\bibinfo {title} {Motion of spin polariton bullets in semiconductor microcavities},\ }\href {https://doi.org/10.1103/PhysRevLett.107.146402} {\bibfield  {journal} {\bibinfo  {journal} {Phys. Rev. Lett.}\ }\textbf {\bibinfo {volume} {107}},\ \bibinfo {pages} {146402} (\bibinfo {year} {2011})}\BibitemShut {NoStop}%
\bibitem [{\citenamefont {Gippius}\ \emph {et~al.}(2007)\citenamefont {Gippius}, \citenamefont {Shelykh}, \citenamefont {Solnyshkov}, \citenamefont {Gavrilov}, \citenamefont {Rubo}, \citenamefont {Kavokin}, \citenamefont {Tikhodeev},\ and\ \citenamefont {Malpuech}}]{GippiusPRL2007}%
  \BibitemOpen
  \bibfield  {author} {\bibinfo {author} {\bibfnamefont {N.~A.}\ \bibnamefont {Gippius}}, \bibinfo {author} {\bibfnamefont {I.~A.}\ \bibnamefont {Shelykh}}, \bibinfo {author} {\bibfnamefont {D.~D.}\ \bibnamefont {Solnyshkov}}, \bibinfo {author} {\bibfnamefont {S.~S.}\ \bibnamefont {Gavrilov}}, \bibinfo {author} {\bibfnamefont {Y.~G.}\ \bibnamefont {Rubo}}, \bibinfo {author} {\bibfnamefont {A.~V.}\ \bibnamefont {Kavokin}}, \bibinfo {author} {\bibfnamefont {S.~G.}\ \bibnamefont {Tikhodeev}},\ and\ \bibinfo {author} {\bibfnamefont {G.}~\bibnamefont {Malpuech}},\ }\bibfield  {title} {\bibinfo {title} {Polarization multistability of cavity polaritons},\ }\href {https://doi.org/10.1103/PhysRevLett.98.236401} {\bibfield  {journal} {\bibinfo  {journal} {Phys. Rev. Lett.}\ }\textbf {\bibinfo {volume} {98}},\ \bibinfo {pages} {236401} (\bibinfo {year} {2007})}\BibitemShut {NoStop}%
\bibitem [{\citenamefont {Liew}\ \emph {et~al.}(2008)\citenamefont {Liew}, \citenamefont {Kavokin},\ and\ \citenamefont {Shelykh}}]{LiewPRL2008}%
  \BibitemOpen
  \bibfield  {author} {\bibinfo {author} {\bibfnamefont {T.~C.~H.}\ \bibnamefont {Liew}}, \bibinfo {author} {\bibfnamefont {A.~V.}\ \bibnamefont {Kavokin}},\ and\ \bibinfo {author} {\bibfnamefont {I.~A.}\ \bibnamefont {Shelykh}},\ }\bibfield  {title} {\bibinfo {title} {Optical circuits based on polariton neurons in semiconductor microcavities},\ }\href {https://doi.org/10.1103/PhysRevLett.101.016402} {\bibfield  {journal} {\bibinfo  {journal} {Phys. Rev. Lett.}\ }\textbf {\bibinfo {volume} {101}},\ \bibinfo {pages} {016402} (\bibinfo {year} {2008})}\BibitemShut {NoStop}%
\bibitem [{\citenamefont {Li}\ \emph {et~al.}(2024)\citenamefont {Li}, \citenamefont {Chen},\ and\ \citenamefont {\textit{et al.}}}]{LiNP2024}%
  \BibitemOpen
  \bibfield  {author} {\bibinfo {author} {\bibfnamefont {H.}~\bibnamefont {Li}}, \bibinfo {author} {\bibfnamefont {F.}~\bibnamefont {Chen}},\ and\ \bibinfo {author} {\bibfnamefont {H.~J.}\ \bibnamefont {\textit{et al.}}},\ }\bibfield  {title} {\bibinfo {title} {All-optical temporal logic gates in localized exciton polaritons},\ }\href {https://doi.org/10.1038/s41566-024-01483-2} {\bibfield  {journal} {\bibinfo  {journal} {Nat. Photonics}\ }\textbf {\bibinfo {volume} {18}},\ \bibinfo {pages} {864} (\bibinfo {year} {2024})}\BibitemShut {NoStop}%
\bibitem [{\citenamefont {Zhao}\ \emph {et~al.}(2024)\citenamefont {Zhao}, \citenamefont {Fieramosca},\ and\ \citenamefont {\textit{et al.}}}]{ZhaoNC2024}%
  \BibitemOpen
  \bibfield  {author} {\bibinfo {author} {\bibfnamefont {J.}~\bibnamefont {Zhao}}, \bibinfo {author} {\bibfnamefont {A.}~\bibnamefont {Fieramosca}},\ and\ \bibinfo {author} {\bibfnamefont {R.~B.}\ \bibnamefont {\textit{et al.}}},\ }\bibfield  {title} {\bibinfo {title} {Room-temperature polariton spin switches based on van der waals superlattices},\ }\href {https://doi.org/10.1038/s41467-024-51612-2} {\bibfield  {journal} {\bibinfo  {journal} {Nat. Commun.}\ }\textbf {\bibinfo {volume} {15}},\ \bibinfo {pages} {7601} (\bibinfo {year} {2024})}\BibitemShut {NoStop}%
\bibitem [{\citenamefont {Opala}\ and\ \citenamefont {Matuszewski}(2023)}]{OpalaOME2023}%
  \BibitemOpen
  \bibfield  {author} {\bibinfo {author} {\bibfnamefont {A.}~\bibnamefont {Opala}}\ and\ \bibinfo {author} {\bibfnamefont {M.}~\bibnamefont {Matuszewski}},\ }\bibfield  {title} {\bibinfo {title} {Harnessing exciton-polaritons for digital computing, neuromorphic computing, and optimization},\ }\href {https://doi.org/10.1364/OME.496985} {\bibfield  {journal} {\bibinfo  {journal} {Opt. Mater. Express}\ }\textbf {\bibinfo {volume} {13}},\ \bibinfo {pages} {2674} (\bibinfo {year} {2023})}\BibitemShut {NoStop}%
\bibitem [{\citenamefont {Laussy}\ \emph {et~al.}(2006)\citenamefont {Laussy}, \citenamefont {Shelykh}, \citenamefont {Malpuech},\ and\ \citenamefont {Kavokin}}]{LaussyPRB2006}%
  \BibitemOpen
  \bibfield  {author} {\bibinfo {author} {\bibfnamefont {F.~P.}\ \bibnamefont {Laussy}}, \bibinfo {author} {\bibfnamefont {I.~A.}\ \bibnamefont {Shelykh}}, \bibinfo {author} {\bibfnamefont {G.}~\bibnamefont {Malpuech}},\ and\ \bibinfo {author} {\bibfnamefont {A.~V.}\ \bibnamefont {Kavokin}},\ }\bibfield  {title} {\bibinfo {title} {Effects of bose-einstein condensation of exciton polaritons in microcavities on the polarization of emitted light},\ }\href {https://doi.org/10.1103/PhysRevB.73.035315} {\bibfield  {journal} {\bibinfo  {journal} {Phys. Rev. B}\ }\textbf {\bibinfo {volume} {73}},\ \bibinfo {pages} {035315} (\bibinfo {year} {2006})}\BibitemShut {NoStop}%
\bibitem [{\citenamefont {van Kampen}(2007)}]{vanKampen2007}%
  \BibitemOpen
  \bibfield  {author} {\bibinfo {author} {\bibfnamefont {N.~G.}\ \bibnamefont {van Kampen}},\ }\href {https://www.elsevier.com/books/stochastic-processes-in-physics-and-chemistry/van-kampen/978-0-444-52965-7} {\emph {\bibinfo {title} {Stochastic Processes in Physics and Chemistry}}},\ \bibinfo {edition} {3rd}\ ed.\ (\bibinfo  {publisher} {Elsevier},\ \bibinfo {year} {2007})\BibitemShut {NoStop}%
\bibitem [{\citenamefont {Rubo}\ \emph {et~al.}(2003)\citenamefont {Rubo}, \citenamefont {Laussy}, \citenamefont {Malpuech}, \citenamefont {Kavokin},\ and\ \citenamefont {Bigenwald}}]{RuboPRL2003}%
  \BibitemOpen
  \bibfield  {author} {\bibinfo {author} {\bibfnamefont {Y.~G.}\ \bibnamefont {Rubo}}, \bibinfo {author} {\bibfnamefont {F.~P.}\ \bibnamefont {Laussy}}, \bibinfo {author} {\bibfnamefont {G.}~\bibnamefont {Malpuech}}, \bibinfo {author} {\bibfnamefont {A.~V.}\ \bibnamefont {Kavokin}},\ and\ \bibinfo {author} {\bibfnamefont {P.}~\bibnamefont {Bigenwald}},\ }\bibfield  {title} {\bibinfo {title} {Dynamical theory of polariton amplifiers},\ }\href {https://doi.org/10.1103/PhysRevLett.91.156403} {\bibfield  {journal} {\bibinfo  {journal} {Phys. Rev. Lett.}\ }\textbf {\bibinfo {volume} {91}},\ \bibinfo {pages} {156403} (\bibinfo {year} {2003})}\BibitemShut {NoStop}%
\bibitem [{\citenamefont {Rubo}(2004)}]{RuboPSSA2004}%
  \BibitemOpen
  \bibfield  {author} {\bibinfo {author} {\bibfnamefont {Y.~G.}\ \bibnamefont {Rubo}},\ }\bibfield  {title} {\bibinfo {title} {Kinetics of bose condensation of cavity polaritons},\ }\href {https://doi.org/10.1002/pssa.200304061} {\bibfield  {journal} {\bibinfo  {journal} {Phys. Status Solidi A}\ }\textbf {\bibinfo {volume} {201}},\ \bibinfo {pages} {641} (\bibinfo {year} {2004})}\BibitemShut {NoStop}%
\bibitem [{\citenamefont {Wouters}(2007)}]{WoutersPRB2007}%
  \BibitemOpen
  \bibfield  {author} {\bibinfo {author} {\bibfnamefont {M.}~\bibnamefont {Wouters}},\ }\bibfield  {title} {\bibinfo {title} {Resonant polariton-polariton scattering in semiconductor microcavities},\ }\href {https://doi.org/10.1103/PhysRevB.76.045319} {\bibfield  {journal} {\bibinfo  {journal} {Phys. Rev. B}\ }\textbf {\bibinfo {volume} {76}},\ \bibinfo {pages} {045319} (\bibinfo {year} {2007})}\BibitemShut {NoStop}%
\bibitem [{\citenamefont {Mandel}\ and\ \citenamefont {Wolf}(1995)}]{MandelWolf1995}%
  \BibitemOpen
  \bibfield  {author} {\bibinfo {author} {\bibfnamefont {L.}~\bibnamefont {Mandel}}\ and\ \bibinfo {author} {\bibfnamefont {E.}~\bibnamefont {Wolf}},\ }\href {https://doi.org/10.1017/CBO9781139644105} {\emph {\bibinfo {title} {Optical Coherence and Quantum Optics}}}\ (\bibinfo  {publisher} {Cambridge University Press},\ \bibinfo {year} {1995})\BibitemShut {NoStop}%
\bibitem [{Not()}]{Note1}%
  \BibitemOpen
  \href@noop {} {}\bibinfo {note} {We use the terms "single-photon-seeded" and "quantum-initiated" to emphasize that a single photon can trigger the stimulated build-up of a macroscopic condensate whose polarization then persists. Our model is semiclassical (stochastic Gross–Pitaevskii with classical noise) and does not implement a complete quantum-memory protocol for arbitrary unknown quantum states, nor does it assess entanglement preservation. Accordingly, we refrain from claiming a full quantum memory and focus on polarization alignment and retention enabled by bosonic stimulation.}\BibitemShut {Stop}%
\bibitem [{\citenamefont {Brune}\ \emph {et~al.}(2025)\citenamefont {Brune}, \citenamefont {Rozas}, \citenamefont {West}, \citenamefont {Baldwin}, \citenamefont {Pfeiffer}, \citenamefont {Beaumariage}, \citenamefont {Alnatah}, \citenamefont {Snoke},\ and\ \citenamefont {A{\ss}mann}}]{BruneCommMat2025}%
  \BibitemOpen
  \bibfield  {author} {\bibinfo {author} {\bibfnamefont {Y.}~\bibnamefont {Brune}}, \bibinfo {author} {\bibfnamefont {E.}~\bibnamefont {Rozas}}, \bibinfo {author} {\bibfnamefont {K.}~\bibnamefont {West}}, \bibinfo {author} {\bibfnamefont {K.}~\bibnamefont {Baldwin}}, \bibinfo {author} {\bibfnamefont {L.~N.}\ \bibnamefont {Pfeiffer}}, \bibinfo {author} {\bibfnamefont {J.}~\bibnamefont {Beaumariage}}, \bibinfo {author} {\bibfnamefont {H.}~\bibnamefont {Alnatah}}, \bibinfo {author} {\bibfnamefont {D.~W.}\ \bibnamefont {Snoke}},\ and\ \bibinfo {author} {\bibfnamefont {M.}~\bibnamefont {A{\ss}mann}},\ }\bibfield  {title} {\bibinfo {title} {Quantum coherence of a long-lifetime exciton-polariton condensate},\ }\href {https://doi.org/10.1038/s43246-025-00848-6} {\bibfield  {journal} {\bibinfo  {journal} {Commun. Mater.}\ }\textbf {\bibinfo {volume} {6}},\ \bibinfo {pages} {123} (\bibinfo {year} {2025})}\BibitemShut {NoStop}%
\bibitem [{\citenamefont {Sedov}\ and\ \citenamefont {Kavokin}(2025)}]{SedovLight2025}%
  \BibitemOpen
  \bibfield  {author} {\bibinfo {author} {\bibfnamefont {E.}~\bibnamefont {Sedov}}\ and\ \bibinfo {author} {\bibfnamefont {A.}~\bibnamefont {Kavokin}},\ }\bibfield  {title} {\bibinfo {title} {Polariton lattices as binarized neuromorphic networks},\ }\href {https://doi.org/10.1038/s41377-024-01719-4} {\bibfield  {journal} {\bibinfo  {journal} {Light: Sci. Appl.}\ }\textbf {\bibinfo {volume} {14}},\ \bibinfo {pages} {52} (\bibinfo {year} {2025})}\BibitemShut {NoStop}%
\bibitem [{\citenamefont {Cuevas}\ \emph {et~al.}(2018)\citenamefont {Cuevas}, \citenamefont {Silva}, \citenamefont {Roch},\ and\ \citenamefont {et~al.}}]{CuevasSciAdv2018}%
  \BibitemOpen
  \bibfield  {author} {\bibinfo {author} {\bibfnamefont {{\'A}.}~\bibnamefont {Cuevas}}, \bibinfo {author} {\bibfnamefont {F.}~\bibnamefont {Silva}}, \bibinfo {author} {\bibfnamefont {J.~G.}\ \bibnamefont {Roch}},\ and\ \bibinfo {author} {\bibnamefont {et~al.}},\ }\bibfield  {title} {\bibinfo {title} {First observation of the quantized exciton-polariton field and effect of interactions on a single polariton},\ }\href {https://doi.org/10.1126/sciadv.aao6814} {\bibfield  {journal} {\bibinfo  {journal} {Sci. Adv.}\ }\textbf {\bibinfo {volume} {4}},\ \bibinfo {pages} {eaao6814} (\bibinfo {year} {2018})}\BibitemShut {NoStop}%
\end{thebibliography}%

\end{document}